\documentclass[journal=jpcafh,manuscript=article,layout=onecolumn]{achemso}

\usepackage{bm}
\usepackage{amsmath}
\usepackage{amssymb}
\usepackage{txfonts}
\usepackage{graphicx}
\usepackage{xspace}
\usepackage{float}
\usepackage{units}
\usepackage{color}

\usepackage{tikz}
\usepackage{pgfplots}
\pgfplotsset{compat=1.18}
\usepackage{booktabs}
\usepackage{multirow}
\usepackage{subcaption}
\usepackage[normalem]{ulem}
\usepackage{tabularx}

\SectionNumbersOn

\author{Gilberto A. Alou Angulo}
\email{gilberto.alou@univ-lille.fr} 

\affiliation{Univ. Lille, CNRS, UMR 8523 – PhLAM – Physique des Lasers Atomes et Molécules, F-59000 Lille, France}

\author{Alejandro Rivero Santamar\'{i}a}
\email{alejandro.rivero@univ-lille.fr}
\affiliation{Univ. Lille, CNRS, UMR 8523 – PhLAM – Physique des Lasers Atomes et Molécules, F-59000 Lille, France}

\author{Céline Toubin}
\email{celine.toubin@univ-lille.fr}
\affiliation{Univ. Lille, CNRS, UMR 8523 – PhLAM – Physique des Lasers Atomes et Molécules, F-59000 Lille, France}

\author{Maurice Monnerville}
\email{maurice.monnerville@univ-lille.fr}
\affiliation{Univ. Lille, CNRS, UMR 8523 – PhLAM – Physique des Lasers Atomes et Molécules, F-59000 Lille, France}

\title{Ab Initio Molecular Dynamics calculations on NO oxidation over oxygen functionalized Highly Oriented Pyrolytic Graphite} 

\begin{document}

\maketitle

\begin{abstract}

The oxidation of NO molecules on epoxy-functionalized highly oriented pyrolytic graphite, thermalized at 300 K, was studied by means of ab initio molecular dynamics (AIMD) calculations. Four collision energies and two different orientations were analyzed where the reaction, adsorption, and scattering probabilities were computed. Our results reveal that NO$_2$ formation can occur even at the lowest collision energy investigated (0.025 eV), approximately equivalent to room temperature (300 K), which agrees qualitatively with the experimental results. This underscores the influence of dynamics on the NO oxidation process, since this oxidation barrier was previously theoretically estimated to be about 0.1 eV at 0 K, which is four times higher than our lowest collision energy. Additionally, we obtained angular and energy distributions of the products under selected simulation conditions. Scattered NO molecules show low specular reflection, lose half of their initial translational energy, and remain vibrationally cold with minimal rotational excitation. Furthermore, a statistical analysis of all reactive trajectories, focusing on configurations at specific reaction moments, elucidated the structural requirements for the reaction to occur under dynamic conditions. Finally, this study demonstrates the potential of oxygen-doped carbon surfaces for the conversion of NO to NO$_2$.

\end{abstract}

\section{Introduction}
\label{sec:introduction}

Nitrogen monoxide (NO) is primarily generated from the combustion of fossil fuels and its presence in the atmosphere is of major concern due to its contribution to environmental problems such as photochemical smog and acid rain\cite{de_vries_impacts_2021}. Therefore, controlling and reducing NO emissions from anthropogenic activities is crucial for environmental protection and sustainability, where selective catalytic reduction (SCR) is the most widely used process for the de-NO$_x$ of fuel gases, despite several drawbacks, such as high reaction temperatures ($>300^{\circ}$C) and unreacted reducing agents \cite{Han_SCRreview_2019}. In this sense, catalytic oxidation of NO to nitrogen dioxide (NO$_2$) at ambient temperatures presents a promising solution to control NO emissions, as NO$_2$ can be converted further to nitric acid in the presence of water\cite{zhao_review_2021}. For this reaction, carbon-based materials have been regarded as good catalysts in recent decades\cite{shen_catalytic_2016}, but although several investigations have been carried out \cite{mochida_oxidation_1994, mochida_removal_2000, mochida_no_2000}, the reaction mechanism remains debated. 

In this regard, an understanding of the mechanism of the adsorption and oxidation of NO on activated carbon at the atomic level is necessary, and its implications go beyond providing insight for the development of more effective carbon-base catalysts to reduce NO emissions. In the context of atmospheric chemistry, it could provide also new insights in the formation process of secondary pollutants compounds such as nitrous acid (HONO) and NO$_2$ by the oxidation of NO on soot particles in the atmosphere. Since soot, resulting from the incomplete combustion of hydrocarbon fuels, is typically formed by aggregated carbonaceous spherules having graphitic structures, graphite surfaces are usually used as proxies of soot particles \cite{xi_review_2021, stanmore_oxidation_2001, saini_carbon_2021, li_microphysical_2024}.

In previous experimental studies of the NO oxidation mechanism over carbon surfaces, Ahmed et al.\cite{ahmed_catalytic_1993} suggested that NO is first oxidized by oxygen to form NO$_2$ in the gas phase and then NO$_2$ gets adsorbed on the carbon surface. However, the homogeneous oxidation of NO in the gas phase is too slow to account for the conversion that was further observed on activated carbon \cite{sousa_no_2012}. Adapa et al.\cite{adapa_catalytic_2006} proposed that NO oxidation is catalyzed through the Langmuir-Hinshelwood (L-H) and Eley-Rideal (E-R) mechanisms. In their study, in the L-H mechanism, dissociated oxygen activated by carbon reacts with adsorbed NO while in the E-R mechanism, it is assumed that adsorbed NO reacts with gaseous O$_2$ and produces adsorbed NO$_2$. Mochida et al.\cite{mochida_oxidation_1994, mochida_removal_2000, mochida_no_2000} also proposed an E-R mechanism in which NO is first adsorbed, then oxidized, forming adsorbed NO$_2$ before being released to the gas phase. However, Sousa et al.\cite{sousa_no_2012} based on previous literature \cite{KONG19961027, kaneko_analytical_1997, stanmore_oxidation_2008}, assume that NO from the gas phase reacts with chemisorbed oxygen, forming adsorbed NO$_2$. It was assumed that very little NO can be physically adsorbed, since it is considered a supercritical gas at ambient temperature. Indeed, the presence of functional oxygen groups in graphite significantly increases the adsorption probability of NO on its surface\cite{suzuki_study_1994} and its subsequent oxidation \cite{sousa_catalytic_2011}.

Using density functional theory (DFT) calculations, Tang and Cao \cite{tang_adsorption_2011} investigated the interaction of NO$_x$ ($x=1, 2, 3$) with graphene and graphene oxides (GO) surfaces presenting some functional groups such as hydroxyl, epoxy, and carbonyl. They report that the adsorption of nitrogen oxides on graphene oxides is generally stronger than that on graphene, but the oxidation of NO was not taken into account. Hou et al.\cite{hou_adsorption_2015} and Cen et al.\cite{cen_oxidation_2015} using DFT studied the effects of hydroxyl and epoxy groups on the adsorption of NO and its further oxidation to NO$_2$. They found that NO can be oxidized by epoxy groups with a barrier of 0.1 eV. However, the kinetic mechanism of the catalytic oxidation of the NO over the epoxy group was not investigated.

In this context, the understanding of the interaction between NO and carbon-based compounds is fundamental to decrypt the complexity of these catalytic oxidation processes. To this aim, we perform ab initio molecular dynamics (AIMD) calculations of the collisions of gaseous NO molecules an an epoxy-functionalized highly oriented pyrolytic graphite surface. The use of AIMD has the main advantage that is not necessary to precalculate a potential energy surface (PES), the energies and forces being determined on the fly during the dynamics at the density functional theory level. Also, a better description of the molecule-surface interaction, including molecule-surface energy exchange and temperature effects, can be effectively achieved.

The paper is organized as follows. Section \ref{sec:computational details} describes the theoretical method and the practical implementation of the AIMD simulations on the present system. The methodology is applied in Sec.~\ref{sec:results} in order to determine the reaction probabilities, the scattering angle distribution and the energy transfer. Finally, a summary of the results and the main conclusions of this work are provided in Sec. \ref{sec:Conclusions}.

\section{Computational details}
\label{sec:computational details}

Ab initio molecular dynamics (AIMD) calculations were performed with the DFT based Vienna Ab Initio Simulation Package (VASP)\cite{kresse_efficiency_1996, kresse_efficient_1996} using the Perdew-Burke-Ernzerhof (PBE)\cite{perdew_generalized_1996} exchange-correlation functional, together with the DFT-D3(BJ) dispersion correction\cite{grimme_consistent_2010, grimme_effect_2011} that amends the inadequacy of PBE to describe van der Waals (vdW) interactions. The ionic cores were described with the projector augmented wave (PAW) method\cite{blochl_projector_1994} implemented in VASP\cite{kresse_ultrasoft_1999}. A gaussian smearing of 0.05 eV was used for the electron occupancies and the plane wave expansion was cut at a kinetic energy of 400 eV.

Since NO and NO$_2$ are open-shell molecules, we tested both spin-polarized and non-spin-polarized calculations to validate the system parameters. Both calculation strategies led to the same results regarding the minimum energy adsorption configurations, showing a minimal influence of the NO$_x$ magnetic properties on the system under study, as previously reported by Hou et al. \cite{hou_adsorption_2015}.

\subsection{Gas-Surface model} 

The lattice parameters calculated for bulk graphite using a (16$\times$16$\times$8) Monkhorst-Pack sampling\cite{monkhorst_special_1976} of the Brillouin zone are $a=b=$~2.47~{\AA} and $c=$~6.70~{\AA}. The graphite (0001) surface was then described by a three layer (4$\times$4) supercell (96 atoms of C) separated from its periodic image by 20~{\AA} of vacuum along the surface normal. In all calculations performed with this supercell, the Brillouin zone was sampled with a (5$\times$5$\times$1) Monkhorst-Pack mesh. After surface relaxation the interlayer distance remains as 3.37~{\AA}. These values are in agreement with experimental \cite{baskin_lattice_1955} and previous theoretical DFT-D3(BJ)\cite{rego_comparative_2015, lebedeva_comparison_2017, rivero_santamaria_ab_2019} results (3.34 and 3.41~{\AA}, respectively). 

To adequately describe the oxidized HOPG (O-HOPG) the binding energy $E_{b}$ of the O atom on HOPG was calculated on the basis of the usual definition: $E_{b} = E_{slab + O(a)} - (E_{O(g)} + E_{slab})$ where $E_{slab + O(g)}$, $E_{O(a)}$ and $E_{slab}$ correspond to the optimized total energy of the system with the oxygen adsorbed on the surface, the energy of an isolated oxygen atom in the gas phase and the energy of pristine graphite, respectively. The geometry optimization was performed using a convergence criterion for the ion step of 0.01~eV/{\AA} and 10$^{-6}$ for the electronic step. The O atom was initially positioned at a distance of 1.5~{\AA} from the surface at three different locations on the graphite surface (bridge, hollow and top). After relaxation, the bridge position was found to be the most stable, with $E_{b}$ = -2.19 eV, and bond distances for C-O and C-C of 1.46~{\AA} and 1.50 {\AA} respectively. These results are consistent with prior theoretical\cite{mehmood_adsorption_2013} and experimental\cite{larciprete_atomic_2012} studies. 

The description of the adsorbate was also benchmarked with respect to gas phase values. For a NO molecule at a distance of 10 {\AA} and aligned parallel to the O-HOPG surface, the obtained dissociation energy (7.17~eV) and the equilibrium bond length (1.17~{\AA}) agree reasonably well with the experimental values (binding energy of 6.55~eV and bond length of 1.15~{\AA}, respectively)\cite{huber_constants_1979, bach_bond_2021}. In addition, using the same calculation strategy, the adsorption energies of NO on the epoxy group of the O-HOPG surface and of NO$_2$ on the pristine HOPG surface were calculated. The obtained values, -0.17 eV for NO and -0.26 eV for NO$_2$, are in agreement with previous theoretical and experimental findings \cite{so_no2_1990,tang_adsorption_2011,hou_adsorption_2015} confirming the reliability of the chosen approach to model NO and NO$_2$ interaction with graphite surfaces.

\subsection{Molecular Dynamics setup}

\begin{figure}[ht]
\includegraphics [width=0.5\columnwidth,angle=0.] {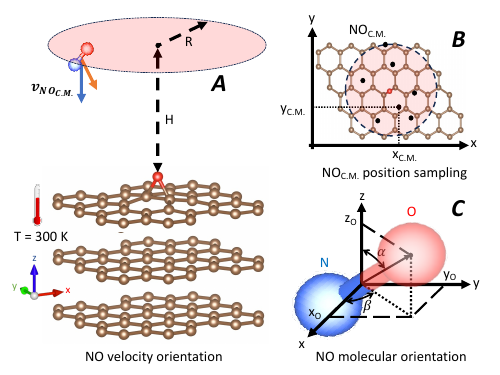 }
\caption{Coordinate system and sketch of the initial conditions used in the AIMD simulations of the NO molecule impinging over the O-HOPG surface where the height $ H $ and the radius $ R $ are 7.5~\AA ~and 5~\AA ~respectively. The values $ x_{C.M.} $, $ y_{C.M.}$, were randomly selected within the disk of raidus $ R $ (black dots shown in panel B). $ \alpha $ and $\ \beta $ were randomly generated.}
\label{fig:scheme}
\end{figure}

The collision between an incoming NO molecule and the O-HOPG surface, thermalized at 300 K, was simulated using quasi-classical trajectories using constant-energy AIMD calculations. For each trajectory, the center of mass of the colliding NO molecule (NO$_{CM}$) was randomly positioned inside a disk with a radius R of 5~{\AA}, centered at the adsorbed oxygen atom and located 9~{\AA} away from the oxidized graphite surface, as represented in panels A and B of Figure \ref{fig:scheme}. The internal orientation ($\theta$, $\varphi$) of the NO molecule, represented in the panel C of the Figure \ref{fig:scheme}, was also randomly selected, and the initial internuclear distance and internal momentum of NO were adapted to semi-classical conditions corresponding to the ro-vibrational state ($v=0$, $j=0$) of the NO molecule. Four different NO incidence energies were studied: 0.025 eV (290 K), 0.050 eV (580 K), 0.100 eV (1160 K) and 0.300 eV (3481 K). These selected collision energies cover a wide range, from high collision energies (0.3 eV) to those equivalent to room temperature (0.025 eV), including a collision energy that matches the value of the energy barrier (0.1 eV) reported in a previous theoretical study\cite{hou_adsorption_2015}. The orientation of the NO$_{CM}$ velocity vector was defined in two different ways (panel A, Figure \ref{fig:scheme}). First, with fixed polar and azimuthal incidence angles $\theta_{i}=0^\circ$ and $\varphi_{i}=0^\circ$, corresponding to normal NO incidence conditions. Second, by adapting its initial orientation ($\theta_{i},\varphi_{i}$) to target the oxygen atom adsorbed on the surface, in order to increase the probability of NO-O collision, this will be referred to as oriented incidence conditions. 

The initial O-HOPG surface configurations were obtained by running a canonical AIMD calculation for 14 ps with a time step of 1 fs, using the Nosé-Hoover thermostat implemented in VASP. This procedure guarantees an initial surface temperature in the simulations equivalent to room temperature (300 K), emulating atmospheric conditions. Then, following a standard procedure in gas-surface dynamics simulations \cite{gros_hydrogen_2007, nattino_effect_2012, nattino_ab_2014, kolb_communication_2016, novko_energy_2017, zhou_ab_2017, zhou_communication_2018, fuchsel_reactive_2019, rivero_santamaria_ab_2019}, one snapshot (positions and velocities) of the previously thermalized O-HOPG surface was randomly selected for each trajectory to perform constant energy AIMD calculations. This procedure is a reasonable approximation for the short simulation times of interest (4 ps).

 Finally, under these conditions, 300 AIMD trajectories per initial incidence energy were performed in the two different orientations of the NO$_{CM}$ velocity vector (normal incidence and oriented incidence) which represents in total 2400 trajectories. The maximum propagation time was 4 ps with a timestep of 1 fs. Four possible exit channels were observed: i) two reactive : NO$_2$ scattered or NO$_2$ adsorbed and ii) two non-reactive: NO scattered or NO adsorbed. Since the dissociation energy of NO on an O-HOPG surface obtained by our calculations is 4.2 eV, the dissociative adsorption channel of NO is closed under our simulation conditions. The molecules NO or NO$_2$ were considered desorbed when their center of mass reached a distance of 7~{\AA} above the surface and adsorbed if after 4~ps the condition just mentioned was not reached. The total energy was almost conserved along the trajectories with a standard deviation of $\sim$6 meV.

 Reflected trajectories were discriminated according to the difference between the initial NO molecule incidence angle $\theta_i$ and the final product (NO or NO$_2$) scattering angle $\theta_s$, computed as $\Delta\theta = \theta_s - \theta_i$. Angular distributions of the desorbed molecules were calculated using constant intervals of 9$^\circ$. The energy distributions of the molecules along the dynamics were obtained using the standard semi-classical determination method for the translational, vibrational and rotational energies \cite{billing_dynamics_2000}.

\section{Results and Discussion}
\label{sec:results}

\subsection{Reaction probabilities and reaction mechanism}
Table \ref{tab:oriented_react_prob} summarizes the probability of occurrence for the observed exit channels at the incidence energies studied, considering two different orientations of the initial velocity vector (normal and oriented incidences). Total probabilities are computed by adding adsorbed and scattered molecules for both reactive and non-reactive channels. The results show a dependence with respect to the initial orientation of the NO molecule velocity vector. For normal incidence, the reaction probability (NO$_2$ formation) is low, in the range 7.3~\% - 9.3~\%, and does not show dependence on the initial NO kinetic energy. On the other hand, when the NO incidence is oriented to target the O atom on the HOPG surface (in parentheses), a clear increase in the reaction probability is observed, rising from 13.6~\% to 26.3~\% as the initial kinetic energy of the incident NO molecule increases from 0.025 to 0.300 eV. 

\begin{table*}[ht]
    \centering
    \caption{Reaction and scattering probabilities of the 300 trajectories at each incidence energy for normal incidence and oriented incidence (in parentheses)}
    \label{tab:oriented_react_prob}
    \begin{tabular}{ccccccc}
        \toprule
        \multirow{2}{*}{E$_k$ (eV)} & \multicolumn{3}{c}{NO$_2$ (reactive channel)} & \multicolumn{3}{c}{NO (non-reactive channel)}\\ \cmidrule(lr){2-4} \cmidrule(lr){5-7}
        & Scattered (\%) & Adsorbed (\%) & Total (\%) & Scattered (\%) & Adsorbed (\%) & Total (\%)\\ 
        \midrule
        0.025 & 4.33 (6.70) & 5.00 (7.00) & 9.33 (13.6) & 5.67 (1.60) & 85.0 (84.7) & 90.7 (86.4)\\ 
        0.050 & 6.33 (10.7) & 1.00 (6.30) & 7.33 (17.0) & 11.0 (7.60) & 81.7 (75.3) & 92.7 (83.0)\\ 
        0.100 & 6.33 (12.3) & 2.33 (7.30) & 8.67 (19.6) & 23.7 (15.3) & 67.7 (65.0) & 91.3 (80.4)\\ 
        0.300 & 7.33 (20.0) & 1.67 (6.30) & 9.00 (26.3) & 57.0 (35.7) & 34.0 (38.0) & 91.0 (73.7)\\ 
        \bottomrule
    \end{tabular}
\end{table*}

The distinct behavior observed can be explained by analyzing the initial position of the center of mass (c.m.) of the impinging NO molecule. Our calculations demonstrate that, irrespective of the incidence energy, the probability of reaction is significantly enhanced when the c.m. of the NO molecule is initially located in the $xy$ plane inside a disk of radius $2.5~\AA$  centered on the epoxy surface group. More than 89~\% and 72~\% of reactive events take place in this region for normal and oriented incidences, respectively. This condition is crucial for normal incidence; if the c.m. of NO is outside of this disk, the oxidation reaction is very unlikely to occur. In fact, our calculations indicate that under these conditions, NO$_2$ is only formed when the NO molecule bounces at least once on the surface, i.e., in an indirect manner. For cases of oriented incidence, we also observe a decrease in the reaction probability when the position of the NO molecule c.m in the $xy$ plane is far from the O atom on the surface. However, this decrease is compensated by the initial orientation of the NO velocity vector, which favors the direct encounter between the NO molecule and the oxygen atom of the HOPG surface. The higher the incident energy of NO, the greater the probability of NO$_2$ formation. If the distance, in the $xy$ plane, between the center of mass of NO (c.m.) and the oxygen atom on the surface is greater than $2.5~\AA$, the molecule must approach with a favorable orientation and sufficient energy to trigger the reaction. In contrast, when the NO center of mass position in the $xy$ plane falls within a radius of $2.5~\AA$ around the epoxy group on the surface, the reaction becomes substantially more probable and is independent of the incidence energy. 

Interestingly, our calculations show that the reaction (NO$_2$ formation) can occur even at the lowest collision energy investigated (0.025 eV), roughly equivalent to room temperature (300 K). These findings qualitatively align with earlier experimental observations, which indicate a high likelihood of NO oxidation on oxidized graphite surfaces at room temperature. \cite{adapa_catalytic_2006, mochida_no_2000, mochida_oxidation_1994}. Quantitatively, the reaction probabilities observed in our simulation are relatively low, amounting to less than  20 \%, in contrast to the experimental probabilities\cite{mochida_no_2000} ($\sim$ 80~ \%). However, we must take into account that performing a direct comparison of the simulation results with the experimental data is challenging because of the characteristics of the simulations (simulation time, number of incident molecules, number of epoxy groups on the HOPG surfaces). In this sense, our results indicate the right trend, if an encounter between a NO molecule and an epoxy group on the surface of HOPG occurs, the possibility of the reaction is not null, it is expected that the multiplication of these encounters over a long period of time may result in a higher reaction probability as the experimental observations show. 

From a theoretical perspective, Hou et al.\cite{hou_adsorption_2015}, using the nudge elastic band method (NEB) based on DFT calculations, identified an energy barrier of 0.1 eV (1160 K) for NO oxidation in graphene oxides. This barrier remains too significant to observe reactivity at room temperature, a contradiction to the outcomes of our simulations. However, it should be noted that the calculations performed by  Hou et al.\cite{hou_adsorption_2015} were not conducted under dynamical conditions; in NEB calculations, the surface is considered at 0 K. Thermal effects, taken into account in our AIMD trajectories, contribute to enhance the likelihood of the oxidation reaction of NO with the epoxy group of graphite. The influence of surface temperature effectively lowers the reaction barrier, which, under room temperature conditions, can be confidently asserted to be below 0.025 eV. As the temperature of graphite rises, for instance at temperatures exceeding 370 K \cite{larciprete_dual_2011, sun_surface_2011, xiaowei_effect_2004}, the bonds formed between oxygen and graphite are weakened, occasionally leading to the desorption of the oxygen \cite{larciprete_atomic_2012}. 

\begin{figure}[ht]
    \centering
    \includegraphics[width=0.5\columnwidth,angle=0.]{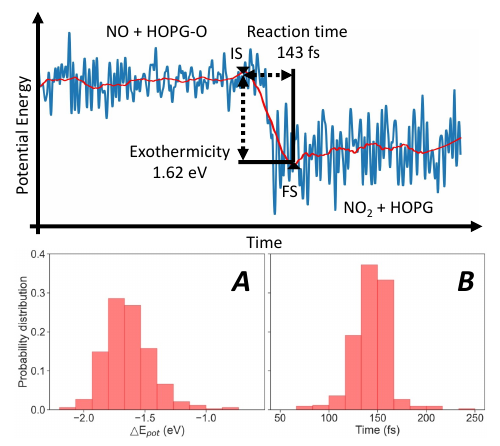}
    \caption{Top Panel: Time evolution of the total potential energy along a typical reactive trajectory. IS designates the initial state, marking the point when the potential energy of the system starts to decrease, while FS indicates the final state, where the potential energy of the system stabilizes and NO$_2$ is formed. Panel A: Probability distribution of the energy released during the reaction. Panel B: Probability distribution of the reaction time computed as ($\Delta t = t_{FS} - t_{IS}$). Both distributions are normalized to all the reactive trajectories from the four initial kinetic energies (0.025, 0.05, 0.1, 0.3 eV) and the two investigated incidences (normal and oriented). The size of the bins are 0.13 eV and 16.6 fs respectively.}
    \label{fig:React_mechanism}
\end{figure}

The key to understanding the reactivity in this system lies in the exothermic nature of the NO oxidation reaction on the O-HOPG surface. The top panel of Figure \ref{fig:React_mechanism} shows the typical evolution of the potential energy along a reactive trajectory. As can be seen, there is a sudden decrease in potential energy during the reaction. This decline occurs at a specific moment, termed the initial state (IS), marking when the system attains the necessary structural and energetic conditions for the reaction to commence. Subsequently, a notable amount of potential energy is liberated within the system, converting into kinetic energy until the potential energy stabilizes once more. At this point, henceforth identified as the final state (FS), the reaction has already occurred and the NO$_2$ molecule has been formed. The availability of kinetic energy plays a pivotal role in elucidating reactivity; once the reaction conditions are met, the kinetic energy in the system is sufficient to break the bonds between the O and C atoms on the surface, providing enough translational energy for the formed NO$_2$ molecule to move away from the surface.

Notably, the average duration of the reaction process, from the initial state (IS) to the final state (FS), is independent of both, the incident energy and the type of incidence, with an average value well centered around 143 fs as showed in panel B of Figure \ref{fig:React_mechanism}. The obtained reaction time indicates a rapid oxidation process, suggesting a direct reaction pathway.

Deepening on the exothermicity of the reaction, the panel A in the Figure \ref{fig:React_mechanism} represents the distribution of the potential energy differences obtained between the FS and the IS ($\Delta E = E_{FS} - E_{IS}$) across all reactive trajectories. This distribution spans from -2.2 eV to -0.8 eV, with a peak at -1.6 eV, suggesting the most probable potential energy difference between reactants and products. This average value is significant for understanding the energetics of the reaction and provides insight into the energy accessible to break the C-O bonds in the epoxy group and facilitate the removal of the nascent NO$_2$ molecule from the surface of HOPG under our simulation conditions. In a previous DFT study\cite{hou_adsorption_2015}, the exothermicity of this process on a very similar system (graphene oxides) was estimated to be -2.38 eV. Although this value is larger than the one obtained in our AIMD simulations, these  differences can be attributed to the fact that our calculation takes into account the surface temperature and the dynamic conditions of the reaction, whereas the NEB calculation assumes a surface temperature of 0 K. In summary, the reaction's exothermicity, in conjunction with surface temperature, significantly aids the breaking of the C-O bonds within the epoxy group, leading to NO$_2$ formation even at relatively low NO incidence energies. 

\begin{figure}[ht]
    \centering
    \includegraphics[width=\columnwidth,angle=0.]{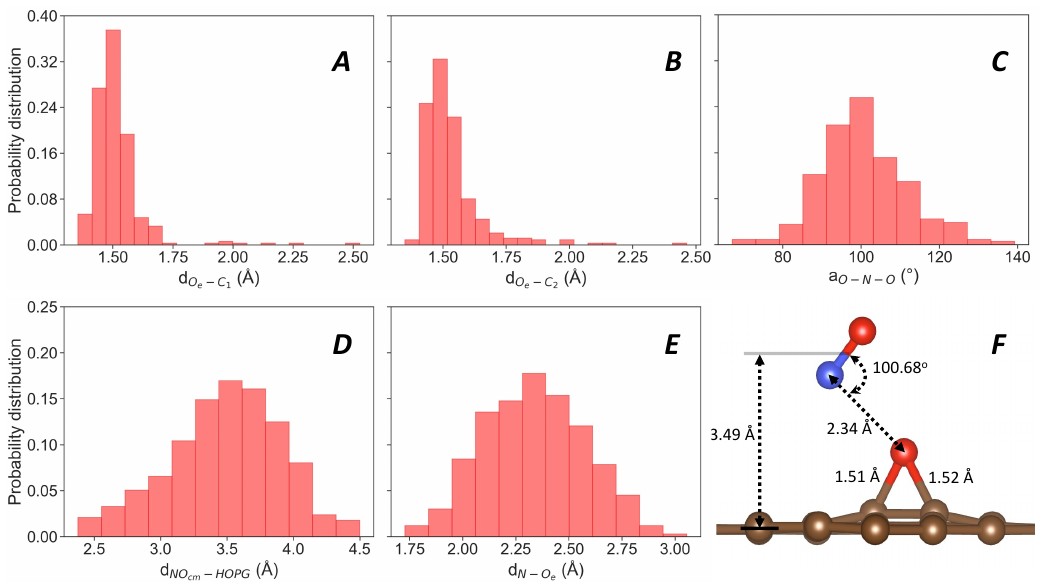}
    \caption{Probability distribution of the geometrical configuration  between the NO and the epoxy group at the initial state (IS). All distributions are normalized to all the reactive trajectories from the four initial kinetic energies (0.025, 0.05, 0.1, 0.3 eV) and the two investigated incidences (normal and oriented). The size of the bins are 0.06 Å, 0.06 Å, 6.0$^\circ$, 0.17 Å and 0.11 Å respectively.}
    \label{fig:IS_distributions}
\end{figure}

To understand the geometry requirements for the reaction to occur under our simulation conditions, we performed a statistical analysis of the configurations across all reactive trajectories at the moment identified as the initial state (IS). For this purpose, average values have been calculated for the following parameters: distance between the N atom and the O atom of the epoxy group ($d_{N-O_{e}}$), distance between the center of mass (c.m.) of the NO molecule and the surface of HOPG along the normal ($d_{NO_{cm}-HOPG}$), angle defined by the N atom, the O atom of the NO molecule and the O atom of the epoxy group ($a_{O-N-O_e}$), and the distance between the O atom and the carbons in the epoxy group ($d_{O_{e}-C_{1}}$, $d_{O_{e}-C_{2}}$). The corresponding distributions are shown in the panels A, B, C, D and E of Figure \ref{fig:IS_distributions}. The configuration characterized by the average values is depicted in panel F of Figure \ref{fig:IS_distributions}. As can be seen, the c.m. of the NO molecule is located 3.49~\AA\ from the O-HOPG surface, with the N atom facing the epoxy group at a distance of 2.34~\AA. The formed O-N-O angle is 100.68$^\circ$ and the distance between the adsorbed oxygen and the carbon atoms are 1.51~\AA\ and 1.52~\AA\ respectively.

When we compare the average configuration obtained in our calculations for the initial state (IS) with the one obtained by Hou et al.\cite{hou_adsorption_2015} using the NEB methodology for the transition state (TS) of NO oxidation on graphene oxide, some interesting correspondences appear. In their reported values, the N atom is located at 2.17~{\AA} (2.34 $\pm$ 0.24~{\AA} in this work) from the adsorbed oxygen, which is at 3.58~{\AA} (3.49 $\pm$ 0.42~{\AA} in this work) ahead of the graphene surface. The distance between the carbon atoms bonded to the adsorbed oxygen and the oxygen atom itself are 1.66~{\AA} (1.51 $\pm$ 0.12~{\AA} in this work) and 1.47~{\AA} (1.52 $\pm$ 0.12~{\AA} in this work), respectively. These comparisons indicate that the IS averaged configuration is very close to the previously predicted TS \cite{hou_adsorption_2015}, highlighting that the geometric conditions necessary for the reaction to occur are met. It confirms that our AIMD simulations follow the same reaction pathway. In summary, to efficiently form NO$_2$ from the NO oxidation reaction over room temperature graphitic surfaces, the NO molecules should be at a distance of $\sim$ 3.50~{\AA} from the surface and $\sim$ 2.30~{\AA} from the surface oxygen atom, oriented with the N atom facing the epoxy group, and forming an O-N-O angle of $\sim$ 100$^\circ$.

Regarding the non-reactive events, an analysis of the probabilities reported in Table \ref{tab:oriented_react_prob} shows that the NO adsorption probability decreases with increasing incident energy, from $\sim$ 85~\% at 0.025 eV to $\sim$ 36~\% at 0.3 eV, regardless of the initial orientation. This result indicates, as expected, that the increase in NO incidence energy makes the trapping of the NO molecule on the surface less probable and points out the minor role of the incidence conditions on the adsorption probability in non-reactive collisions.

\subsection{Products angular distributions}

\begin{figure}[H]
    \centering
    \includegraphics[width=0.5\columnwidth,angle=0.]{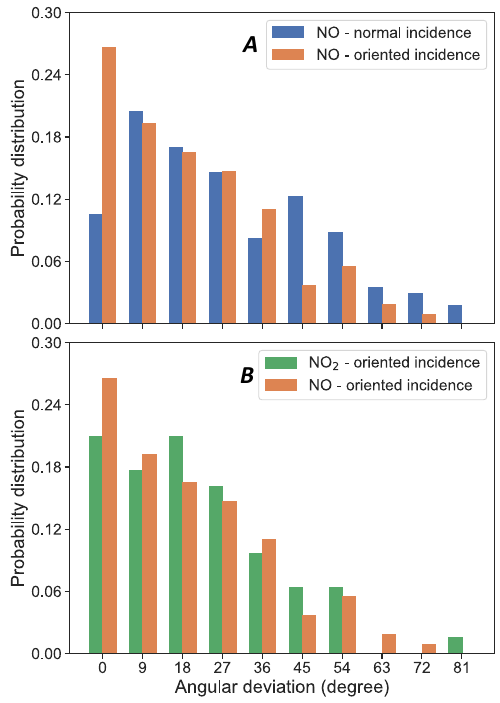}
    \caption{Comparative analysis of the distributions of the angular deviations $\Delta\theta$ after reflection from the oxidized graphite surface at an initial kinetic energy of 0.3 eV. The distributions are normalized to the total number of scattered molecules and the bin size is 9$^\circ$. (A) for NO scattered molecules with initial normal and oriented incidences; (B) for desorbed NO and NO$_2$ molecules for initial NO with oriented incidence.}
    \label{fig:delta_ang_dist}
\end{figure}

Recognizing the complex interplay between the admolecules and the surface, we conducted an examination of the scattering dynamics of NO and NO$_2$ molecules upon reflection from the thermally equilibrated O-HOPG surface. This analysis concentrated on trajectories featuring the highest investigated initial incidence energy of the NO molecule, set at 0.3 eV, for both normal and oriented incidences. This particular initial incidence energy was selected because it corresponds to the highest percentage of NO and NO$_2$ molecules desorbed from the surface in both directions of incidence, thus providing a more robust statistical data set. Considering the initial conditions used in this work, where the incidence angle is 0$^\circ$ for NO normal incidence and varies in a range between 0$^\circ$ and 18$^\circ$ for NO oriented incidence, we have evaluated the difference between the initial NO molecule incidence angle $\theta_i$ and the final product (NO or NO$_2$) scattering angle $\theta_s$, computed as $\Delta\theta = \theta_s - \theta_i$. The probability distribution of this deviation angle is presented in Figure \ref{fig:delta_ang_dist}. Note that for normal incidence, this angular distribution corresponds directly to the NO scattering distribution since the initial angle $\theta_{i}$ is equal to 0.

For the non-reactive NO scattering trajectories, our results indicate some differences in the angular distributions obtained under the two different incidence conditions. For the oriented incidence, the distribution represented in the panel A of Figure~\ref{fig:delta_ang_dist} peaks at smaller angles, indicating a higher probability for specular reflection ($\theta_s = \theta_i$) than for the normal incidence conditions. On the other hand, the normal incidence angular distribution exhibits a more spread-out shape, indicating a higher probability of molecules with large scattering angles ($\theta_s >> \theta_i=0$). The observed differences might be attributed to two main factors. First, in the oriented incidence cases, the direct interaction between the incident NO molecule and the epoxy group favors elastic scattering, leading to an increase in the number of molecules scattered relatively close to the specular reflection. Second, in normal incidence cases, the z component of the velocity vector of the molecule tends to experience the most substantial change during collision with a thermalized surface, favoring inelastic scattering and resulting in an increase in the number of molecules scattered at larger angles. 

For reactive trajectories, where NO$_2$ molecules are desorbed after the oxidation of NO, the panel B of Figure \ref{fig:delta_ang_dist} presents the $\Delta\theta$ distribution of NO$_2$ molecules compared with the previously obtained $\Delta\theta$ distribution of NO molecules, but only for the oriented incidence case. Due to the limited number of NO$_2$ molecules produced (low reaction probability) under normal incidence conditions, there is insufficient statistical data to represent and compare the corresponding distributions. Our results show a similar shape for both distributions, indicating that the deviation angle of the scattered NO molecules or formed NO$_2$ molecules, in case of oriented incidence, follows the same behavior for reactive and non-reactive events. This observation could be ascribed to the fast NO oxidation process, indicated by the previously estimated reaction time of 143 fs, implying a direct reaction pathway. Consequently, the nascent NO$_2$ molecules pursue a trajectory similar to that of NO scattered molecules in the non-reactive scenarios.

\subsection{Products energy distributions}

Translational, vibrational and rotational energy distributions were calculated for NO scattered molecules at the highest incidence energy (0.3 eV) for both normal and oriented incidences. Additionally, the translational energy distribution was obtained for the NO$_2$ desorbed molecules at the same incidence energy, but only for the oriented incidence conditions. 

\begin{figure}[H]
    \centering
    \includegraphics[width=0.5\columnwidth,angle=0.]{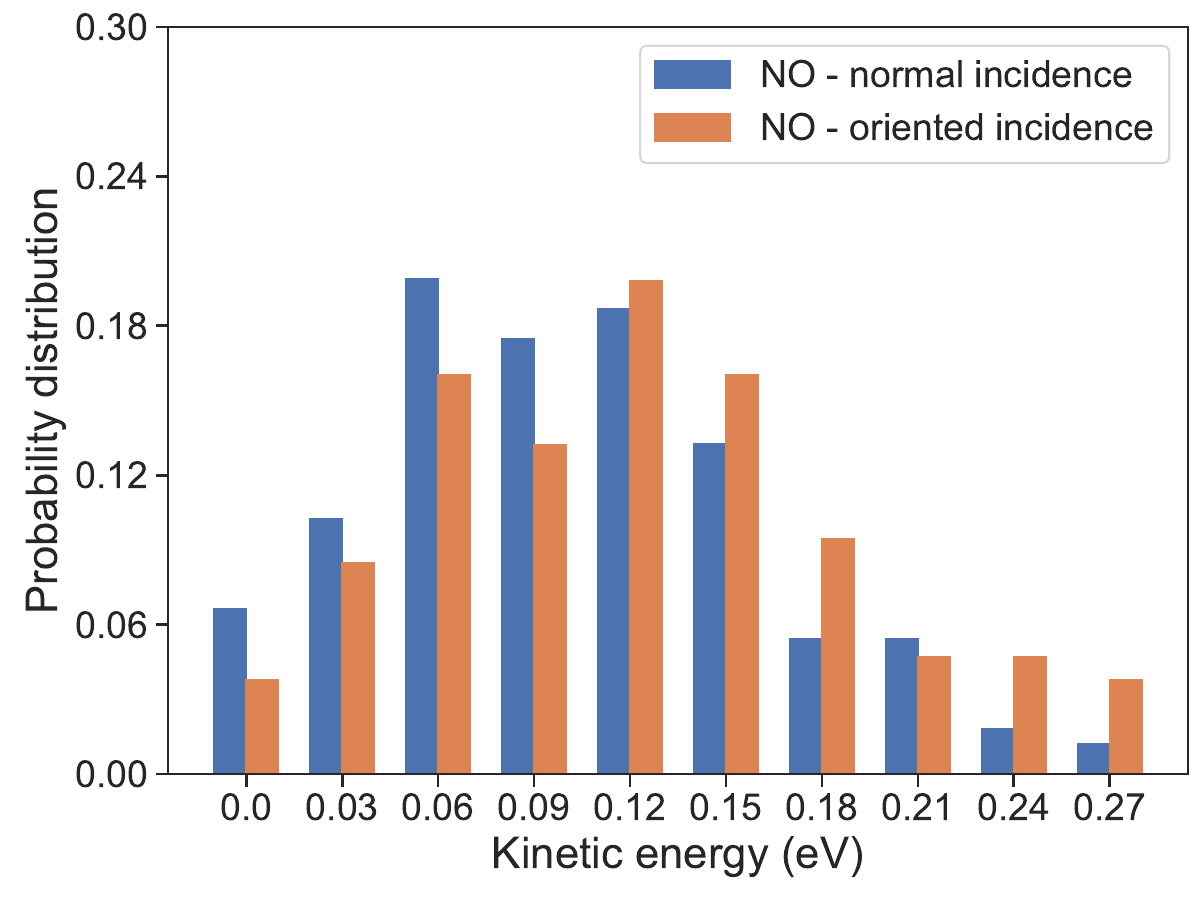}
    \caption{Translational energy distribution of NO molecules scattered with an initial kinetic energy of 0.3 eV at normal and oriented incidences. The distributions are normalized to the total number of scattered NO molecules and the bin size is 0.03 eV.}
    \label{fig:Ek_NO}
\end{figure}

The translational energy distributions of NO scattered molecules for both initial incidences are shown in Figure \ref{fig:Ek_NO}. As can be seen, both distributions have a similar bell shape, but slight differences can still be appreciated. The normal incidence distribution peaks at lower kinetic energy values ($\sim$ 0.06 eV), with a distribution mean of 0.12 eV, indicating an average translational energy loss of 60.0~\%. On the other hand, for those molecules whose incidence was oriented, the distribution peaks at higher kinetic energy values ($\sim$ 0.12 eV), with a distribution mean of 0.14 eV, indicating an average translational energy loss of 53.3~\%. These results are in agreement with the scattering patterns previously analyzed, as a higher angular deviation at normal incidence indicates a higher translational energy loss compared to oriented incidence conditions. The main factor contributing to the observed differences between the oriented and normal incidences, regarding both the deviation angle and translational energy distribution, seems to be the direct NO-O interaction. This interaction results in a stronger molecule-surface repulsion effect, which is more pronounced under oriented incidence conditions where the NO-O collision is forced to occur. This strong interaction could slightly reduce the loss of kinetic energy during the molecule-surface collision process. 

Furthermore, our results are also in good agreement with previous experimental observations on similar systems. Häger et al. \cite{Hager_1985} found that at low pyrographite surface temperatures ($<$ 300 K), the NO scattered has a significantly smaller mean velocity, approximately half that of the incoming molecule. Additionally, in a recent study, Greenwood and Koehler\cite{greenwood_nitric_2021} investigated the scattering of NO molecules in gold-supported room temperature graphene and observed a translational energy loss of the scattered NO molecule of approximately 80~\%. In general, our AIMD simulations indicate that for a slightly different system, with the presence of an epoxy group on the graphite surface, there is also quite efficient kinetic energy transfer from the molecule to the surface at room temperature, resulting in a translational energy loss of the incoming molecule greater than 50~\% after the collision.

\begin{figure}[H]
    \centering
    \includegraphics[width=0.5\columnwidth,angle=0.]{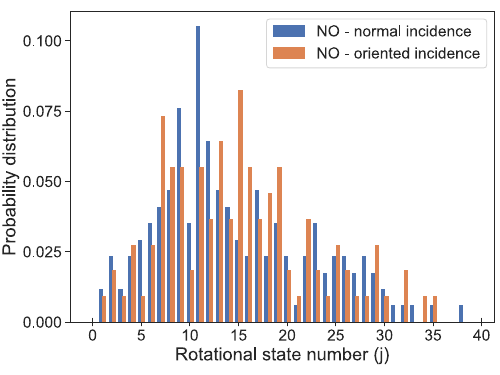}
    \caption{Rotational distribution of scattered NO molecules with 0.3 eV of initial kinetic energy at normal and oriented incidence. The distributions are normalized to the total number of scattered NO molecules and the bin size is 1.}
    \label{fig:j_NO}
\end{figure}

Concerning the ro-vibrational state of the scattered NO molecules, our simulations show, in agreement with previous observation on similar systems\cite{greenwood_nitric_2021, Hager_1985}, that no vibrational excitation occurs after the collision and that a small amount of energy is channeled into the rotational mode. Figure \ref{fig:j_NO} shows the rotational energy distribution of the scattered NO with an initial kinetic energy of 0.3 eV at both incidence conditions. Once again, both distributions show a similar bell shape with slight differences in the maximum peaks. For normal incidence, the most probable rotational quantum number for the NO desorbed molecules is $j = 12$, while for oriented incidence, it is $j = 15$. These rotational quantum numbers represent rotational molecular energies of 0.03 eV and 0.05 eV respectively, indicating that only a small amount of the total energy initially available for the NO molecule (0.3 eV in translational energy + 0.1 eV in vibrational energy) was channeled to the molecule rotation after the collision with the surface. Indeed, as observed recently by Greenwood et al.\cite{greenwood_nitric_2021} who studied NO scattering off graphene using surface-velocity map imaging, our simulations indicate that in NO scattering on an O-HOPG surface, most of the collision energy is transferred into the collective motion of the carbon atoms in the topmost layer of the graphite surface and is not channeled from translational to rotational modes of the molecules.

\begin{figure}[H]
    \centering
    \includegraphics[width=0.5\columnwidth,angle=0.]{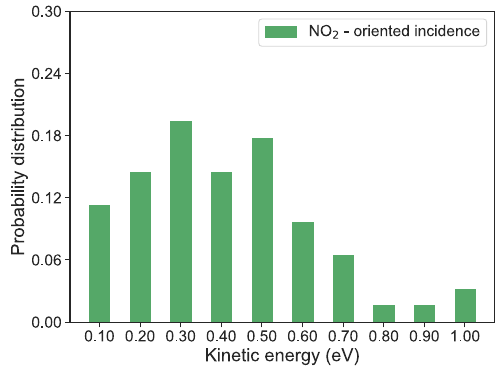}
    \caption{Translational energy distribution of NO$_2$ molecules scattered for an NO initial kinetic energy of 0.3 eV at oriented incidence. The distribution is normalized to the total number of scattered NO$_2$ molecules and the bin size is 0.1 eV.}
    \label{fig:Ek_NO2}
\end{figure}

Finally, Figure \ref{fig:Ek_NO2} shows the translational energy distribution of the NO$_2$ desorbed molecules at the highest incidence energy of 0.3 eV for the oriented incidence conditions. As can be seen, the distribution presents a bimodal shape, with one peak located at 0.3 eV, at the initial incidence energy, and another peak located at 0.5 eV, above the initial incidence energy. The mean value of the NO$_2$ kinetic energy is 0.36 eV, indicating, in contrast to previous results for scattered NO, a small increase in the translational energy of the NO$_2$ molecule of 20~\% relative to the initial kinetic energy of the incoming NO (0.3 eV). This increase in the NO$_2$ molecule’s translational energy is directly linked to the exothermicity of the reaction, which was estimated at approximately 1.62 eV under our molecular dynamics conditions. The available energy resulting from the reactive process is first used to break the C-O bond of the epoxy group. Once the NO$_2$ molecule is formed, the desorbing molecules use part of this energy to escape from the the HOPG surface. The remaining energy is then channeled into the translational and ro-vibrational modes of the nascent NO$_2$ molecule. In this regard, a complete analysis of the energetic characteristics of the NO$_2$ desorbed molecules, including the rotational and vibrational state populations, is reserved for a later publication.

\section{Conclusions}
\label{sec:Conclusions}

In summary, we performed ab initio molecular dynamics simulations to study the oxidation of NO molecules on oxygen-functionalized highly oriented pyrolytic graphite, thermalized at 300 K. A total of 2400 AIMD trajectories were analyzed, with 300 trajectories for each incidence energy (0.025 eV, 0.05 eV, 0.1 eV and 0.3 eV) at two different orientations of the incoming NO molecule (normal to the graphite and oriented to guarantee a collision with the oxygen atom of the epoxy group on the graphite surface). The reaction, adsorption, and scattering probabilities were computed. The angular and energy distributions of the products were also obtained under selected simulation conditions.

Our results reveal that NO$_2$ formation can occur even at the lowest collision energy investigated (0.025 eV), approximately equivalent to room temperature (300 K). This therefore highlights the impact of the dynamics on the NO oxidation process, since previous studies by Hou et al.\cite{hou_adsorption_2015} and Cen et al.\cite{cen_oxidation_2015} had estimated this oxidation barrier at 0 K to be around 0.1 eV, i.e. four times greater than our lowest collision energy of 0.025 eV. Moreover, this result aligns qualitatively with previous experimental observations \cite{mochida_oxidation_1994, mochida_no_2000, dastgheib_no_2020, adapa_catalytic_2006, guo_nano_2021}, which indicate a high likelihood of NO oxidation on oxidized graphite surfaces at room temperature. These findings underscore the efficiency of activated carbonaceous surfaces for facilitating NO conversion and NO$_2$ formation under mild conditions, highlighting their significance for environmental and catalytic applications. 

Concerning the reaction mechanism, our exhaustive analysis shows that NO oxidation is most probable when the direct interaction between the NO molecule and the surface epoxy group is favored. The key to the system’s reactivity is the substantial amount of energy released once the reaction occurs; the exothermicity of the reaction under the simulation conditions was estimated to $\sim$ 1.62 eV. This energy is used to break the surface epoxy group bonds, and in some cases, is enough to overcome a physisorption barrier, allowing the nascent NO$_2$ molecule to escape from the surface. Furthermore, the statistical analysis performed on all the reactive trajectories, selecting configurations at specific reaction moments, allows us to describe the structural requirements for the reaction to occur under dynamical conditions. Remarkably, our results align well with the previous predictions by Hou et al. \cite{hou_adsorption_2015} and Cen et al. \cite{cen_oxidation_2015} regarding the transition state predicted for a very similar system. Moreover, our analysis shows that once the energetic and configurational requirements are reached, the reaction occurs quickly ($\sim$ 143 fs) following a direct reaction mechanism.

We have also performed an analysis of the angular distributions of the NO scattered and NO$_2$ desorbed molecules at selected initial conditions. The results indicate slight differences between the scattered patterns of NO depending on the incidence conditions of the incoming molecule. Specular scattering is favored when the direct interaction with the O atom is forced (oriented incidence). This finding points out a possible feature that may provide clues in future experimental works to identify the interactions that occur between the incident molecule and the surface using scattering angular distributions, and potentially identify the presence of the O atom on the graphite surface. Furthermore, the similarity between the angular distributions of the NO scattered molecules and the NO$_2$ desorbed molecules confirms the direct nature of the NO oxidation reaction on O-HOPG surfaces.

Finally, the energetic characteristics of the NO scattered molecules obtained in our simulations confirm previous findings. There exists an efficient transfer of translational energy from the impinging NO molecule to the graphite surface, with more than 50~\% of the initial NO molecule’s energy being lost in the collision with the graphitic surface. The energy transfer is influenced by the presence of oxygen on the surface, with a slight decrease in energy loss observed when the interaction of the incoming NO molecule occurs directly with the epoxy group present on the surface. Furthermore, a weak rotational excitation is predicted in our simulations for the NO scattered molecules, in agreement with previous theoretical \cite{nyman_surface_1990, greenwood_molecular_2022} and experimental \cite{greenwood_nitric_2021, hager_scattering_2004} results on similar systems. Regarding the translational energy characteristics of the nascent NO$_2$ molecules, our findings indicate a slight increase in the available kinetic energy compared to the incoming NO molecule. In this regard, a complete analysis of the energetic characteristics of the NO$_2$ desorbed molecules, including the rotational and vibrational state populations, is reserved for a later publication.

All in all, our AIMD simulation results shed light on the NO oxidation mechanism on epoxy-functionalized carbon surfaces, indicating the requirements for the reaction to occur and providing some clues about possible product features that can be tracked in future experimental work to further characterize the reaction mechanism.

\section{Acknowledgments}

The authors acknowledge support from the CaPPA project (Chemical and Physical Properties of the Atmosphere) funded by the French National Research Agency (ANR) through the PIA (Programme d’Investissement d’Avenir) under Contract No. ANR-10-LABX-005. The authors thank the Région Hauts-de-France, the Ministère de l’Enseignement Supérieur et de la Recherche and the European Fund for Regional Economic Development for their financial support to the CPER ECRIN program. The authors acknowledge support from the French national supercomputing facilities (Grant Nos. DARI A0130801859, A0110801859) and from the Centre de Ressources Informatiques (CRI) of the Université de Lille. 

\bibliography{Biblio_NO_HOPG_O}

\end{document}